\documentclass[useamsfonts]{pasj01}

\usepackage{xspace}
\usepackage{graphicx}
\usepackage{url}

\def\ngradeA{15}
\def\ngradeB{36}
\def\ngradeAB{51}
\def\ngradeC{282}
\def\ndata{\sim43,000}
\def\nyatta{1480}
\def\nmodel{4751}
\def\narc{5097}
\def\nred{346}

\def\Sref#1{Section~\ref{#1}\xspace}
\def\Fref#1{Figure~\ref{#1}\xspace}
\def\Tref#1{Table~\ref{#1}\xspace}

\def\pr{{\rm P}}
\def\plens{\pr_{\mathrm{lens}}}

\def\thetae{\theta_{\mathrm{E}}}

\def\chitah{{\sc Chitah}}
\def\arcfinder{{\sc Arcfinder}}
\def\ringfinder{{\sc Ringfinder}}
\def\yattalens{{\sc YattaLens}}
\def\sextractor{{\sc SExtractor}}
\def\emcee{{\sc emcee}}
\def\hscpipe{hscPipe}

\begin{document}

\author{Alessandro~\textsc{Sonnenfeld}\altaffilmark{1}$^{*}$}
\author{James~H.~H.~\textsc{Chan}\altaffilmark{2, 3, 4}}
\author{Yiping~\textsc{Shu}\altaffilmark{5}}
\author{Anupreeta~\textsc{More}\altaffilmark{1}}
\author{Masamune~\textsc{Oguri}\altaffilmark{1, 6, 7}}
\author{Sherry~H.~\textsc{Suyu}\altaffilmark{3, 4, 8}}
\author{Kenneth~C.~\textsc{Wong}\altaffilmark{3, 9}}
\author{Chien-Hsiu~\textsc{Lee}\altaffilmark{10}}
\author{Jean~\textsc{Coupon}\altaffilmark{11}}
\author{Atsunori~\textsc{Yonehara}\altaffilmark{12}}
\author{Adam~S.~\textsc{Bolton}\altaffilmark{13}}
\author{Anton~T.~\textsc{Jaelani}\altaffilmark{14}}
\author{Masayuki~\textsc{Tanaka}\altaffilmark{9}}
\author{Satoshi~\textsc{Miyazaki}\altaffilmark{9, 15}}
\author{Yutaka~\textsc{Komiyama}\altaffilmark{9, 15}}
\altaffiltext{1}{Kavli IPMU (WPI), UTIAS, The University of Tokyo, Kashiwa, Chiba 277-8583, Japan}
\altaffiltext{2}{Department of Physics, National Taiwan University, 10617 Taipei, Taiwan}
\altaffiltext{3}{Institute of Astronomy and Astrophysics, Academia Sinica, P.O.~Box 23-141, Taipei 10617 Taiwan}
\altaffiltext{4}{Max-Planck-Institut f{\"u}r Astrophysik, Karl-Schwarzschild-Str.~1, 85748 Garching, Germany}
\altaffiltext{5}{National Astronomical Observatories, Chinese Academy of
Sciences, 20A Datun Road, Chaoyang District, Beijing 100012,
China}
\altaffiltext{6}{Department of Physics, University of Tokyo, 7-3-1 Hongo, Bunkyo-ku, Tokyo 113-0033, Japan}
\altaffiltext{7}{Research Center for the Early Universe, University of Tokyo, 
Tokyo 113-0033, Japan}
\altaffiltext{8}{Physik-Department, Technische Universit\"at M\"unchen, James-Franck-Stra\ss{}e~1, 85748 Garching, Germany}
\altaffiltext{9}{National Astronomical Observatory of Japan, 2-21-1 Osawa, Mitaka, Tokyo 181-8588, Japan}
\altaffiltext{10}{Subaru Telescope, National Astronomical Observatory of Japan, 650 N Aohoku Pl., Hilo, HI 96720, USA}
\altaffiltext{11}{Department of Astronomy, University of Geneva, ch. d'\'Ecogia 16, 1290 Versoix, Switzerland}
\altaffiltext{12}{Department of Astrophysics and Atmospheric Science, Faculty of Science, Kyoto Sangyo University}
\altaffiltext{13}{National Optical Astronomy Observatory, 950 N. Cherry Ave., Tucson, AZ 85719, USA}
\altaffiltext{14}{Astronomical Institute, Tohoku University, 980-8578, 6-3 Aramaki Aoba, Aoba-ku, Sendai, Miyagi, Japan}
\altaffiltext{15}{SOKENDAI(The Graduate University for Advanced Studies), Mitaka,
 Tokyo, 181-8588, Japan}

\title{Survey of Gravitationally-lensed Objects in HSC Imaging (SuGOHI). I. Automatic search for galaxy-scale strong lenses}

\Received{XXX}

\Accepted{YYY}


\email{alessandro.sonnenfeld@ipmu.jp}

\maketitle

\begin{abstract}
The Hyper Suprime-Cam Subaru Strategic Program (HSC SSP) is an excellent survey for the search for strong lenses, thanks to its area, image quality and depth.
We use three different methods to look for lenses among 43,000 luminous red galaxies from the Baryon Oscillation Spectroscopic Survey (BOSS) sample with photometry from the S16A internal data release of the HSC SSP.
The first method is a newly developed algorithm, named \yattalens, which looks for arc-like features around massive galaxies and then estimates the likelihood of an object being a lens by performing a lens model fit.
The second method, \chitah, is a modeling-based algorithm originally developed to look for lensed quasars.
The third method makes use of spectroscopic data to look for emission lines from objects at a different redshift from that of the main galaxy.
 We find $\ngradeA$ definite lenses, $\ngradeB$ highly probable lenses and $\ngradeC$ possible lenses.
Among the three methods, \yattalens, which was developed specifically for this study, performs best in terms of both completeness and purity.
Nevertheless five highly probable lenses were missed by \yattalens\ but found by the other two methods, indicating that the three methods are highly complementary.
Based on these numbers we expect to find $\sim300$ definite or probable lenses by the end of the HSC SSP.
\end{abstract}

\section{Introduction}

Strong gravitational lensing is a very powerful diagnostic tool for the study of the mass distribution in the Universe.
Strong lensing by galaxies has been used to study a variety of topics in both galaxy evolution and cosmology.
These include properties of the lens galaxies themselves, such as their density profile (\cite{T+K02}, \cite{K+T03}) and its evolution \citep{Ruf++11, Bol++12a, Son++13b}, the distribution of satellites \citep{Mor++09, Veg++12, Nie++14, Hez++16}, the stellar initial mass function (IMF, \cite{Tre++10}, \cite{Son++12}, \cite{Bar++13}, \cite{ORF14}, \cite{Sch++14}) or the mass of the central black hole \citep{Mor++08, Won++15, Tam++15}. In addition, galaxy-scale lenses have been used to study the properties of the lensed background source \citep{Slu++12, Jon++13, Old++17} or for the measurement of cosmological parameters \citep{Suy++13, C+A14, Suy++17, Won++17, Bon++17}.

The number of currently known galaxy-scale strong lenses is a few hundred. 
While this number has allowed us to constrain some {\em average} properties of galaxy structure, such as the mean density profile of massive early-type galaxies (ETGs, \cite{Koo++06}),
we are still limited by statistics once we divide the sample in subsets of different lens properties.
One example is the redshift distribution: the vast majority of known lenses are at a redshift below $z=0.5$, drastically limiting our ability to constrain the evolution of the internal structure of ETGs \citep{Son++15} beyond that point in cosmic history.
Since galaxies are complex systems, it is important to collect lensing measurements for objects covering as wide a range in parameter space as possible.

The most straightforward way to look for gravitational lenses is to explore a large area of sky with sub-arcsecond resolution imaging.
A successful example of such an effort is the Strong Lensing Legacy Survey (SL2S, \cite{Cab++07}).
SL2S was based on the CFHT Legacy Survey (CFHTLS) data, consisting of 170~deg$^2$ with a typical $i-$band seeing of $0.7''$.
CFHTLS data was scanned both with automatic lens finding algorithms and with the contribution from citizen scientists, leading to the discovery of an order of a hundred highly probable lenses \citep{Mor++12, Son++13a, Mor++16}.

The Hyper Suprime-Cam Subaru Strategic Program (HSC SSP, \cite{Aih++17}) provides a natural extension of the SL2S. With its planned 1,400~deg$^2$ coverage in five bands, $0.6''$ seeing and $26.2$~mag depth in $i$-band, it is an excellent survey for the purpose of finding new strong lenses.
In this work we apply three different lens finding methods to a sample of 43,000 massive galaxies selected from the Baryon Oscillation Spectroscopic Survey (BOSS, \cite{SWE09}, \cite{Daw++13}) of the Sloan Digital Sky Survey III (SDSS-III, \citet{Eis++11}) with imaging data from the S16A internal data release of the HSC SSP covering $\sim450$~deg$^2$ in five bands.
Two of the lens finding methods are existing algorithms, while one was developed specifically for this study.
The goal of this study is to explore the potential of HSC data for lens finding purposes, as well as to test the capabilities and limitations of different lens finding algorithms.
We found $\ngradeA$ new definite lenses and $\ngradeB$ highly probable ones.

These 51 newly found lenses and lens candidates form the first sample of the Survey of Gravitationally-lensed Objects in HSC Imaging (SuGOHI).
Since the parent sample of targets used for the search consists of massive galaxies, we refer to the corresponding sample of lenses as the SuGOHI galaxy-scale lens sample, or SuGOHI-g.
In future works we will present a new sample of lenses obtained by looking at clusters of galaxies (SuGOHI-c, More et al. in prep.) as well as a sample of lensed quasars (SuGOHI-q, Chan et al. in prep.).

This paper is organized as follows. In \Sref{sect:data} we present the current HSC SSP data and our target selection based on BOSS. In \Sref{sect:yatta} we introduce our new lens finding algorithm. In \Sref{sect:chitah} and \Sref{sect:emline} we describe two other lens finding methods used.
In \Sref{sect:results} we show the sample of newly found lenses. In \Sref{sect:discuss} we discuss the relative performances of the three lens finders and the properties of the new lenses. We conclude in \Sref{sect:concl}.
All images are oriented with North up and East left.

\section{The Data}\label{sect:data}

\subsection{HSC photometry}

HSC \citep{Miy++12} is a 1.5~deg field of view optical camera recently installed on the Subaru Telescope.
The HSC SSP survey (HSC survey, from here on) is expected to cover a 1,400~deg$^2$ area in five bands ($g$, $r$, $i$, $z$ and $y$) to an $i$-band depth of 26.2 by its completion (see \cite{Aih++17} for more details about the survey).
We use photometric data from the S16A data release, which covers $442$~deg$^2$ in all five bands, 178~deg$^2$ of which to the target depth.
The data is processed with the reduction pipeline \hscpipe\ (Bosch et al. in prep.), a version of the Large Synoptic Survey Telescope stack \citep{Ive++08, Axe++10, Jur++15}.
The median seeing is $0.6''$ in $i$-band and $0.8''$ in $g$-band.
The pixel scale of HSC is $0.168''$.
Although data from the S16A release is not public at the time of writing of this paper, about half of the lens candidates presented in this work are also visible in the public data release 1 (PDR1, \cite{Aih++17}). 

Among the patches of sky imaged by HSC, there are three of the CFHTLS fields. This is important for the development and test of our lens finder because a large number of known lenses have been discovered in CFHTLS data.

\subsection{BOSS spectroscopy and target selection}\label{ssec:targets}

Our strong lens search is lens-based: we select objects with properties typical of lens galaxies and then look for the presence of lensed background sources.
A lens-based search
gives us the advantage of a better control over the selection function of lenses, with the drawback of a loss in completeness.
We select lens galaxy candidates from luminous red galaxies (LRGs) in BOSS. 
The BOSS survey consists of two subsamples of LRGs: the LOWZ and CMASS samples. The main difference between the two samples is mostly the redshift distribution: LOWZ galaxies are mostly at $z < 0.4$ while CMASS galaxies are mostly in the range $0.4 < z < 0.7$.
The number of BOSS galaxies with photometry in all five bands the 2016A data release of HSC is $\ndata$, of which $\sim9,000$ from LOWZ and $\sim34,000$ from CMASS.

There are two reasons for selecting lens galaxy candidates from the BOSS survey. Firstly, BOSS targeted the high mass end of the galaxy population. Since the strong lensing cross section increases with lens mass, more massive galaxies are more likely to be lenses.
Secondly, optical spectroscopy data from BOSS allows us to look for signatures from strongly lensed star forming galaxies in the form of emission lines at a different redshift from the lens.
The detection of emission lines from the background galaxy can add crucial information for the classification of a lens candidate. Moreover, a spectroscopic measurement of the redshift of both the lens and source galaxy allows us to convert angular measurements of the Einstein radius into mass measurements.

\section{A new lens search method}\label{sect:yatta}

We developed a new lens finding algorithm, named \yattalens.
The algorithm consists of several steps, each described in detail below.
The basic idea can be summarized in two key points: 1) \yattalens\ looks for arc-like features around massive galaxies and 2) fits simple lens models to these arcs to assess the likelihood of them being lensed galaxies.
\yattalens\ combines in a novel way the key features of arc detection-based lens finding algorithms, such as \arcfinder\ \citep{Ala06, Mor++12} or \ringfinder\ \citep{Gav++14}, and modeling-based algorithms \citep{Mar++09, Cha++15}.

There are several challenges in the search for galaxy-scale lenses in ground-based imaging data.
First of all, the Einstein radius (roughly the mean distance of lensed images from the center of the lens) of typical lenses is on the order of the half-light radius. This means that lensed images are often blended with the surface brightness of the foreground galaxy, making their detection more complicated.
Secondly, as we will show later, many non-lenses exhibit arc-like features due to the presence of spiral arms or tangentially elongated star forming regions,  increasing the risk of false positive detection.
Thirdly, lens galaxies are often surrounded by satellites, companions, or unrelated objects in close proximity on the line of sight. This complicates the identification of lensed images.
\yattalens\ is designed to tackle these challenges.

\subsection{Lens Light subtraction}

The first step in the search for lensed arcs consists of removing the contribution of the candidate lens galaxy from the image, to facilitate the detection of lensed images.
We do this by fitting an elliptical de Vaucouleurs surface brightness profile \citep{deV48} to the $i$-band data in a small ($3''$ radius) region around the center of the galaxy. 
We choose the $i$-band for two reasons: 1) the image quality of HSC data is best in this band, because of the requirement of using $i$-band images for weak lensing
analysis, the main science driver of the survey, and 2) we expect the lens galaxy to outshine lensed background galaxies at this wavelength, since typical lens systems consist of a massive red galaxy in the foreground and a blue star-forming galaxy in the background.

This step is run for all galaxies in the parent sample, therefore it is important that it is performed in the shortest time possible.
For this reason we describe the light distribution of the lens with a de Vaucouleurs profile, which provides a good descrption of the surface brightness of typical massive galaxies while having a relatively small number of free parameters. In a later step, involving only objects with potential arcs around them, the more general S\'{e}rsic profile is used.
The values of the parameters of the best-fit model are found by running a short Markov Chain Monte Carlo (MCMC) with the Python package \emcee\ \citep{For++13}, which ensures an efficient sampling of the parameter space.

We then take the best fit de Vaucouleurs model and subtract a point spread function (PSF) convolved, rescaled version of it from the $i$-band and all other bands used in the analysis (in our case the $g$-band). 
With this step, we are implicitly assuming that there are no color gradients in the lens galaxy.

Examples of lens-subtracted images are shown in the second column of \Fref{fig:examples1}.

\subsection{Lensed arc identification}\label{ssec:arc}

We use \sextractor\ \citep{B+A96} to look for objects in the lens-subtracted $g$-band image.
By using the $g$-band we expect to detect lensed sources up to redshift $z\sim4$ \citep{Ono++17}.
For each object we consider their position relative to the lens light centroid, their axis ratio and orientation, and the size of their footprint. 

We then apply the following series of conditions to determine whether an object can be a lensed arc or not.
\begin{enumerate}
\item A distance from the lens centroid between 3 and 30 pixels ($0.50'' < R < 5.04''$)
\item A minimum ratio between the major and minor axis of 1.4
\item A maximum difference of 30 degrees between the position angle of the major axis of the object and the tangential to the circle centered on the lens and passing through the object centroid.
\item A minimum angular aperture, defined as the angle subtended by the object from the lens centroid, of 25 degrees.
\item A footprint size between 20 and 500 pixels.
\end{enumerate}
The minimum distance requirement in condition 1, as well as the minimum size requirement in condition 5, makes sure that the candidate arc is not just a residual from a non-perfect subtraction of the lens light. The maximum distance constraint is applied because we do not expect galaxy scale lenses to have Einstein radii larger than $5''$. Although there are lenses with larger Einstein radius, we typically refer to these as group-scale or cluster-scale lenses.
Conditions 2 and 3 are applied to only select tangentially elongated objects.
We add condition 4 to select objects that are elongated not just in terms of axis ratio, but that also describe a sizeable arc around the lens in angular terms. This conditions eliminates small objects far away from the lens that happen to pass the orientation and axis ratio requirement given by condition 2 and 3 but are clearly not strongly lensed.
Finally, we add the maximum size requirement in condition 5 to reject catastrophic failures in the object detection process.
Objects that pass all five conditions are kept as potential arcs.

\subsection{Foreground model}\label{ssec:foreground}

If the object detection process described above returned at least one arc candidate, we proceed with a more accurate fit of the lens light and by modeling potential foreground objects.
We turn back to the $i$-band data and fit a S\'{e}rsic profile \citep{Ser68} to the surface brightness distribution of the main galaxy, this time masking out any objects detected in the $g$-band image by \sextractor.
We then go through non-arc-like objects detected in the $i$-band around the lens, one by one in order of decreasing $i$-band flux, and fit each one with a S\'{e}rsic profile. Each time a new object is fit, the structural parameters of the lens and previously fitted objects (i.e. centroid, effective radius, etc.) are kept fixed, as well as the relative $i$-band amplitude between the lens and any other object fit up to that point. Only the overall amplitude of all previously fit objects is varied. 
This prescription is adopted to reduce computation time.
Only objects within $1.3$ times the distance of the farthest arc candidate of the lens are modeled. Objects farther away are simply masked out.
Foreground objects other than the main galaxy are treated as massless: their constribution to the lens model, to be described in subsection \ref{ssec:lens}, is ignored.

Although most known lenses do not have foreground objects other than the lens galaxy itself in the proximity of lensed images, accounting for the presence of foregrounds is important for the rejection of false positive candidates.
An example of foreground object modeling is shown in the first row of \Fref{fig:examples1}. After the lens light subtraction step, a galaxy is detected North of the lens and is then modeled with a S\'{e}rsic profile. The model of the lens and foreground object, together with a model for the lensed source to be described in subsection \ref{ssec:lens}, is shown in the fourth column.

Up to this point, no color information has been used. Among the objects treated as foreground, there might be counter-images of the main arc. In the next subsection we will illustrate how we can classify images based on their color.

\subsection{Multiple image set candidates}

At this stage of the lens finding process, we have candidate arcs and a model for the light distribution of the lens galaxy and surrounding objects.
These objects can be either foregrounds or multiple images of the arcs.
We use color and position information to distinguish between the two possibilities.

First of all, we apply a color selection to the arc candidates themselves. Since we expect lensed galaxies to be relatively blue we require a $g-i$ color smaller than $2.0$.
Although a $g-i$ color of $2.0$ is not a particularly strong limit, it is sufficient to eliminate a frequent source of contaminants: satellite galaxies with major axis oriented along the tangential direction.
In principle, this color selection criterion could have been applied before the foreground object modeling step, in the interest of computation time. However, having an accurate model for the system is important to measure accurate colors and avoid rejecting good lenses for which the light from the arc happens to be contaminated by a foreground object.
Fluxes in $i$ and $g$ band are measured by summing over pixels in the $g$-band footprint as determined by \sextractor. Although in principle we should match the PSF of the two bands for an accurate color measurement, the effect of differing PSFs is small with respect to the required precision.
More important is the effect of modeling systematics from the subtraction of the light from the lens or foreground objects.
To take this into account we add in quadrature to the statistical uncertainty on the flux of individual pixels a systematic uncertainty equal to 20\% of the contribution from the best fit model of all light components.
 
\begin{figure*}
 \includegraphics[width=\textwidth]{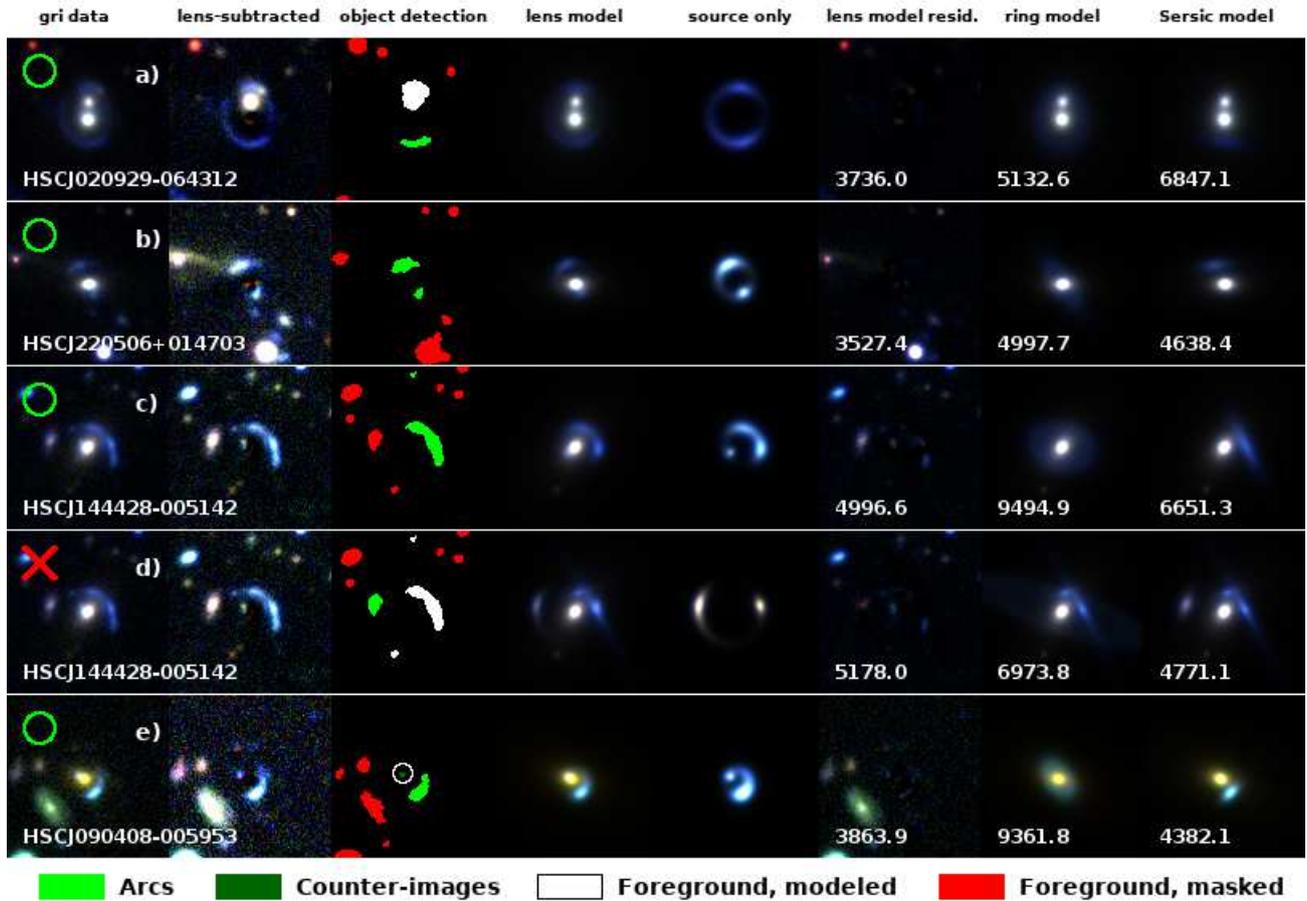}
  \caption{Steps of the lens finding algorithm for a few example lenses and lens candidates. 
The first column shows the $gri$ image of the system. 
The second column contains a lens-subtracted image, displayed with a different contrast to enhance features from lensed objects. 
The third column shows a segmentation map plot of the objects identified by \sextractor. Different colors represent candidate arcs, candidate counter-images to the main arc (only present in the fifth row, circled for better visibility), objects that are modeled as foreground, and foreground objects that are masked out, as specified in the legend. 
The fourth column shows the best fit lens model of the system, including models for foreground objects. 
The fifth column shows the best model of the lensed background galaxy alone, with the contrast set equal to the second column. 
The sixth column shows the residuals between the data and the best-fit lens model.
The seventh and eight columns show the best fit ring and S\'{e}rsic models respectively. 
The numbers at the bottom of the images in column six, seven and eight are the $\chi^2$ of the best fit model for the lens, ring and S\'{e}rsic model respectively.
Circles in the top left corner of each row indicate candidates that passed the selection.
The red cross in the fourth row indicates that the second interpretation of the image configuration for the lens candidate HSCJ144428$-$005142 does not provide a good description of the system, since a model with a single S\'{e}rsic component at the position of the candidate arc gives a better fit.
Even though the images are color-composite, only the $g$-band is used for the modeling of the arc-like features.
The systems shown in rows a), b) and e) are known lenses from the SL2S survey \citep{Mor++12, Son++13a}. The system in rows c) and d) is a new candidate found in HSC data. 
 \label{fig:examples1}
}
\end{figure*}

We go through the blue (according to our color cut) arcs and create sets of arcs and non-arc objects with consistent $g-i$ color, which we interpret as candidate sets of multiple images of the same source. 
We require colors of different objects to be within 2$\sigma$ of each other in order to consider them consistent. 
The minimal multiple image set consists of only one arc. 
If other candidate arcs are present, and if they have consistent $g-i$ color, they are added to the set. An example of such a case is shown in the second row of \Fref{fig:examples1}: two objects satisfying the definition of arc given in subsection \ref{ssec:arc} are found by running \sextractor\ on the lens-subtracted image, shown in bright green in the segmentation map panel (third column). Since they have consistent colors, they are interpreted as multiple images of the same source.

If, on the contrary, there are arc-like objects with a different color than the arc defining the set, and if they are within 1.3 times the distance of the farthest arc from the lens, they are classified as foreground and are fitted with a S\'{e}rsic profile with the same procedure described in subsection \ref{ssec:foreground}. If these arc-like objects are at a farther distance from the lens, they are simply masked out.
An example of this process is shown in the third and fourth row of \Fref{fig:examples1}. For this lens candidate, our algorithm led to the detection of two potential arcs: an extended blue arc to the North-West of the main galaxy and a redder object to the East.
Both objects could be lensed sources, but since they have different colors they cannot be multiple images of the same object. Then, two different interpretations are possible. If the bluer, bigger arc is the lensed source, then the red object must be a foreground. Since this object is too far away from the lens galaxy to affect any further analysis, it is simply masked out. This scenario is shown in the third row: the candidate arc is shown in bright green and objects masked out are shown in red.
In the fourth row, the second interpretation is illustrated: the redder object is considered to be the lensed source, and the bluer elongated feature is modeled as a foreground (marked in white in the segmentation map plot in addition to the main lens galaxy foreground).

Finally, we repeat the same procedure for non-arc-like objects: those with color consistent with the candidate arc(s) are assigned to a multiple image set. Objects with a different color are either modeled as foregrounds or masked out, depending on their distance from the lens galaxy.
In the same system discussed above, a faint object is detected South of the lens. This is masked out in the first interpretation of the image configuration (third row of \Fref{fig:examples1}), but is instead modeled as a foreground in the second interpretation (fourth row), because its distance from the lens is within 1.3 times that of the main arc.

In the fifth row we show an instance of a system in which a non-arc-like object with color consistent to a candidate arc is found. The object is represented in dark green in the segmentation map plot.
This object was modeled as a foreground in the previous step. However, since it is now treated as a counter-image of the arc, it is removed from the model of the foregrounds.

\subsection{Lens model}\label{ssec:lens}

We proceed to fit a lens model to each set of arcs and counter-images.
The lens mass model consists of a singular isothermal ellipsoid (SIE), while the source surface brightness distribution is modeled with a circularly symmetric exponential profile.
The centroid of the lens mass distribution is kept fixed to the value of the lens light centroid.
The free parameters of the model then are: lens angular Einstein radius $\thetae$, axis ratio $q$, position angle PA, source position $(s_x, s_y)$, source effective radius $s_e$, as well as the amplitude of the source, lens and foreground surface brightness components. 
Foreground objects are treated as massless: only their surface brightness profile is modeled.

This is clearly a simplified model compared to what is usually needed to accurately describe strong lens systems, especially with respect to the source surface brightness distribution, a choice that is motivated by the computational time required to fit the data.
However, as we will show later, it is not the absolute quality of a lens model that determines the outcome of the analysis by \yattalens\ but rather its relative quality compared to alternative non-lens models.

We use a modified version of the lens modeling code developed by \citet{Aug++11} to find a maximum-likelihood fit to the data.
We explore the parameter space defined by the six non-linear parameters $\thetae$, $q$, PA, $s_x$, $s_y$ and $s_e$ with an MCMC, where at each step of the chain we optimize for the amplitudes of the surface brightness components.
As for the preliminary lens light subtraction step we use \emcee\ to sample the posterior probability distribution of the model parameters, assuming flat priors on all of them.

This step of the algorithm is particularly slow. In order to minimize the cost in terms of time, we only run a relatively short chain, without ensuring that the MCMC has converged and that the true maximum likelihood model has been found.
In order to maximize the chances of obtaining a good fit it is then important to find a reasonable starting solution.
We use the following prescription to define our starting model. We set the initial mass orientation and axis ratio equal to that of the light distribution. Then, in case only one arc is present, we assume that the Einstein radius is equal to $0.7$ times the distance of the arc from the lens. In case there are two or more arcs, the Einstein radius is set to the mean distance of the arcs from the lens.
Once the lens mass parameters are set, the starting position of the source is found by mapping each arc centroid to the source plane and taking the mean of their position.
We use the $g$-band image to perform the fit. Although the HSC $i$-band data has better image quality, we use the $g$-band to minimize the contamination of the lensed images from the light of the lens galaxy and of foreground objects, which are typically redder than most lensed galaxies.

We then need to assess whether the best-fit lens model is a good description of the data. In principle, we could use the $\chi^2$ to quantify the goodness-of-fit.
In practice, with our simple model it is almost impossible to fit the image of a lens down to noise level, let alone doing it automatically without human intervention.
Therefore, instead of considering absolute values of the $\chi^2$, which have little meaning in this context, we compare the lens model $\chi^2$ with that obtained by fitting alternative models, described below.
In principle, Bayesian evidence can be used to obtain a more accurate comparison between the different models. In practice, a simple $\chi^2$ analysis is sufficient for our purpose and is computationally faster than measuring the evidence.
\subsection{Non-lens models}\label{ssec:nonlens}

The lens candidates that have made it thus far are systems with a relatively blue, tangentially elongated object close to a massive galaxy.
While some of these objects are indeed lenses, most of them are not.
The two main sources of contaminants are 1) spiral arms (we will use the term spiral arm in a broad sense to indicate a mostly tangentially elongated star forming region) and 2) foreground galaxies with high ellipticity.
The non-lens models that we compare against the lens model are designed to describe these two classes of systems.

The first model we consider is a ``ring galaxy'' model.
Its surface brightness profile is defined as follows
\begin{equation}\label{eq:ring}
I(r) = I_0 \left\{\begin{array}{ll} \exp{\{-(r-r_0)/h_o\}} & \rm{for}\,r>r_0 \\
\exp{\{-(r_0 - r)/h_i\}} & \rm{for}\,r<r_0\end{array}\right.
\end{equation}
Here $r$ is the circularized radius of elliptical isophotes,
\begin{equation}
r^2 \equiv qx^2 + y^2/q,
\end{equation}
and $r_0$, $h_i$ and $h_o$ are free parameters describing the radial position of the peak in the surface brightness distribution and an inner and outer scale radius, respectively.
The other free parameters of the model are the ellipticity $q$ and the position angle of the major axis, $\rm{PA}$.

\begin{figure}
 \includegraphics[width=\columnwidth]{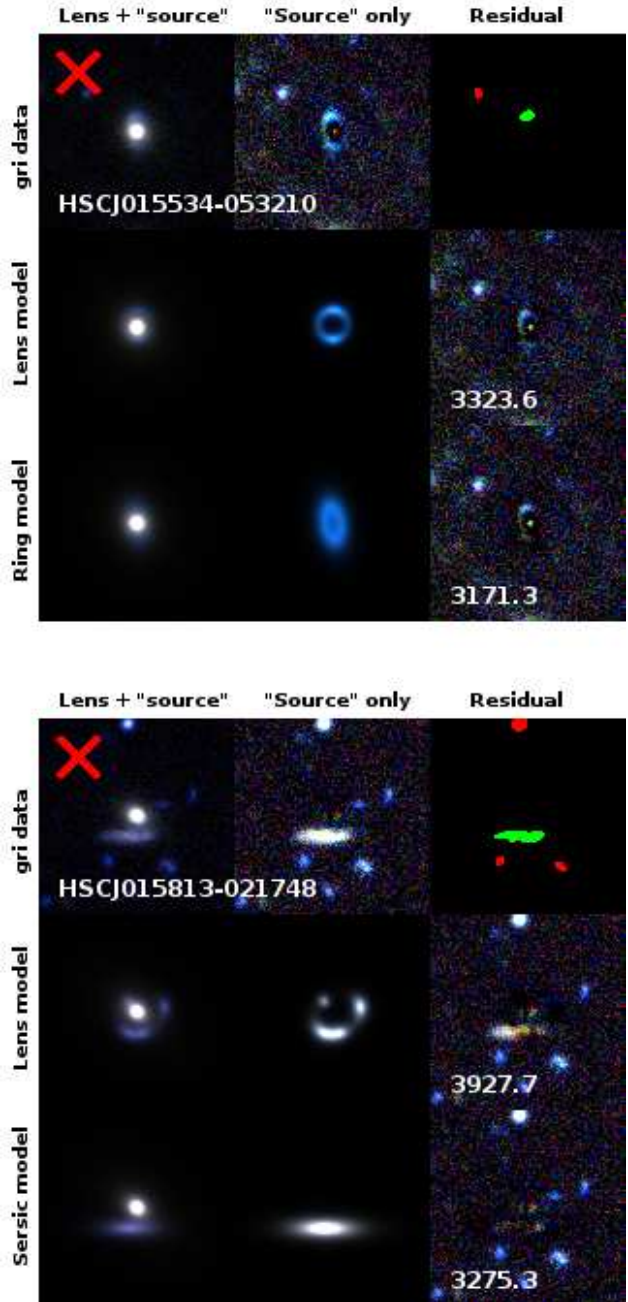}
   \caption{Lens candidates rejected by our algorithm because a ring (top panel) or a S\'{e}rsic (bottom panel) model provides a better fit compared to a lens model. The top row shows a $gri$ image of the system, a lens-subtracted imaged and a segmentation map showing the footprint of the identified candidate arc and that of masked out object.
The second row shows the best-fit lens model of the system, the corresponding source-only image and the residual.
The third row in the top (bottom) panel shows the best-fit ring (S\'{e}rsic) model, a ring-only (S\'{e}rsic-only) image and the residual.
\label{fig:fail}
}
\end{figure}

Different realizations of a ring profile are shown in the seventh column of \Fref{fig:examples1}.
While our lens model cannot produce radially elongated images, due to the upper limit on source size and the choice of a singular isothermal profile, the ring model has the ability of describing relatively large regions with approximately constant surface brightness. For example, for very large values of the inner scale radius $r_i$, the ring profiles reduces to a disk with approximately constant surface brightness within $r_0$. 
Although hardly any galaxy can be well described by this surface brightness profile, the ring model provides a better fit compared to a lens model for most spiral galaxies that happen to be detected by the arc-finding step described in subsection \ref{ssec:arc}, because it can provide a very rough description of their disk.
An example of a galaxy with a candidate arc for which a ring model gives a better fit compared to a lens model is shown in \Fref{fig:fail}.

When fitted to an actual lens, the ring model can hardly provide a reasonable fit. This is because the ring model is elliptically symmetric by construction, while the image configuration of strong lenses is not. The only case in which a ring profile could mimic a lens would be that of a perfectly circular Einstein ring. However, a perfect Einstein ring requires both a perfectly circular mass distribution and a perfect alignment between lens and source, an extremely rare circumstance.

The second non-lens model that we fit is a S\'{e}rsic profile with disky/boxy isophotes. 
Many arc candidates picked by the algorithm are just foreground objects with relatively high ellipticity that happen to be tangentially aligned with respect to the lens.
A S\'{e}rsic profile provides a better fit with respect to a lens model for most of these systems.
One such example is shown in \Fref{fig:fail}, while another instance appears in the fourth row of \Fref{fig:examples1}, where the object interpreted as an arc is better fit by a S\'{e}rsic component.

In conclusion, in the analysis of a system with arc candidates, \yattalens\ first fits a lens model to the $g$-band image and measures the $\chi^2$ of the best-fit model. The $\chi^2$ is defined as
\begin{equation}
\chi^2 = \sum_i \frac{(m_i - d_i)^2}{\sigma_i^2},
\end{equation}
where $m_i$, $d_i$ and $\sigma_i$ are the model flux, measured flux and measurement uncertainty of pixel $i$ and the sum extends over all pixels within $5''$ of the image center, excluding masked out foreground objects.
The $\chi^2$ from the lens model is then compared to the values obtained for the best-fitting ring and S\'{e}rsic models. If the lens model $\chi^2$ is the lowest the candidate is selected. This procedure is carried out for each set of multiple image candidates (i.e. for each image configuration scenario). If in each scenario the lens model provides a worse fit compared to a non-lens model, the candidate is discarded.
\subsection{A summary of the algorithm}

We have described in detail all the steps taken by \yattalens\ to look for strong lenses. Let us summarize the steps of the algorithm. 
\begin{enumerate}
\item Given a massive galaxy, \yattalens\ fits a de Vaucouleurs profile to its $i$-band surface brightness profile, then subtracts the best fit model from the $g$-band image. This step takes a few seconds on a standard machine.
\item Runs \sextractor\ on the lens-subtracted $g$-band image to look for tangentially elongated objects within $\sim5''$. This step takes a fraction of a second.
\item If a candidate arc is found, \yattalens\ proceeds to model any object within a region where potential counter-images of the arc could be, then measures colors of arcs and other objects. This step can take from a few seconds to a few minutes, depending on how many foreground objects are present.
\item If the candidate arc is bluer than $g-i=2$, \yattalens\ makes sets of arcs and images of consistent color. For each multiple image set, any other object with inconsistent color is either modeled as a foreground object or masked out, depending on the ratio between its distance from the lens and the distance of the farthest candidate arc from the lens.
\item For each multiple image set, the $g$-band image is fit with a lens model, a ring model and a S\'{e}rsic model. Each model includes a model of the lens galaxy and possible foreground objects, described as the sum of S\'{e}rsic profiles with fixed relative amplitude. It takes about a minute to run this modeling step.
\item If the lens model provides a better fit, in terms of $\chi^2$, with respect to both the ring and the S\'{e}rsic model, the candidate is selected. Otherwise it is rejected.
\end{enumerate}
In the next subsection we will show how \yattalens\ performs on a sample of known lenses.

\subsection{Tests on known lenses}\label{ssec:known}

The \yattalens\ algorithm has been tested 
on a sample of known lenses that lie in the S16A data release of HSC. These are 16 galaxy-scale lenses from the SL2S survey (SL2SJ021247$-$055552, SL2SJ021411$-$040502, SL2SJ021737$-$051329, SL2SJ022357$-$065142, SL2SJ022511$-$045433, SL2SJ022610$-$042011, SL2SJ022648$-$040610, SL2SJ022708$-$065445, SL2SJ023251$-$040823, SL2SJ023307$-$043838, SL2SJ090407$-$005952, SL2SJ142003$+$523137, SL2SJ220329$+$020518, SL2SJ220506$+$014703, SL2SJ221326$-$000946, SL2SJ222148$+$011542, \cite{Gav++12}, \cite{Son++13a}, \cite{Son++15}), plus the double source plane lens HSCJ142449$-$005322 also known as the ``Eye of Horus'' \citep{Tan++16}.

\begin{figure}
 \includegraphics[width=\columnwidth]{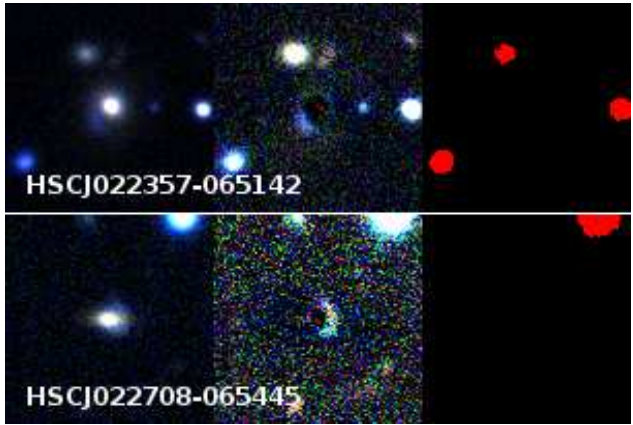}
  \caption{
Known strong lenses from the SL2S survey missed by \yattalens.
The left panels show a color composite $gri$ HSC image of the systems, the middle panels show lens-subtracted images with enhanced contrast and the right panels show the segmentation map obtained by running \sextractor\ on the $g$-band lens-subtracted images. The arcs are not detected.
 \label{fig:missed}
}
\end{figure}

\yattalens\ was able to identify correctly 15 lenses out of 17. The two systems missed by \yattalens\ are SL2SJ022357$-$065142 and SL2SJ022708$-$065445. Both were missed during the arc detection step, the arcs being too faint to be picked out by \sextractor, which we run with a detection threshold of $2\sigma$ above the background.
The images of these two lenses, which were originally discovered using \ringfinder\, are shown in \Fref{fig:missed}.

\begin{figure*}
 \includegraphics[width=\textwidth]{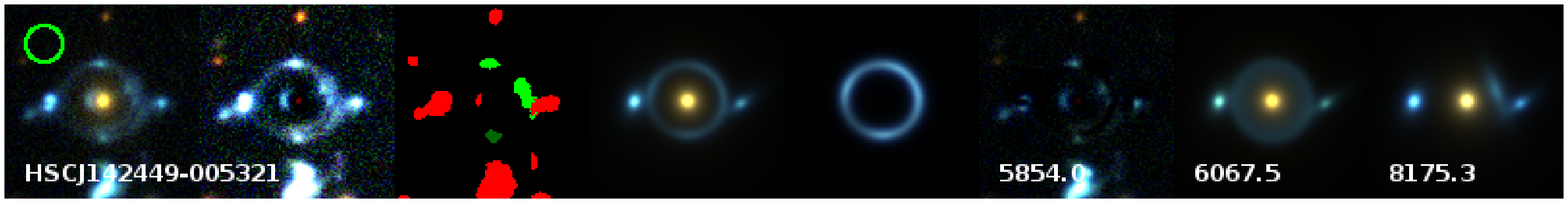}
  \caption{
\yattalens\ analysis of the double source plane lens "The Eye of Horus" \citep{Tan++16}. Panels are the same as \Fref{fig:examples1}. As can be seen from the segmentation map (third panel), \yattalens\ correctly identifies part of the outer ring as an arc. The bright knot on the South part of the outer ring is also correctly identified as a counter-image. The East bright knot however is classified as a contaminant, due to an inaccurate estimate of the color. Nevertheless, the lens model provides a better fit compared to both the ring model (column seven) and the S\'{e}rsic model (column eight) and is the system is correctly classified as lens.
 \label{fig:horus}
}
\end{figure*}

In \Fref{fig:horus} we show \yattalens' analysis on the double source plane lens the Eye of Horus.
Double source plane lenses are very interesting objects because they offer more constraints with respect to typical lenses, and can be used for detailed studies of the mass distribution of the foreground galaxy (e.g. \cite{Son++12}), or to infer cosmological parameters \citep{Gav++08, C+A14, Sch14}.
Since they are also very rare, it is extremely important that lens finding algorithms do recognize them as lenses. This can be challenging for automatic algorithms because the presence of arcs from more than one source complicate their morphology.
Therefore, we made sure when designing \yattalens\ that it could properly classify the Eye of Horus, the only spectroscopically confirmed double source plane lens in the HSC survey, as a lens.

Since most of the SL2S lenses in HSC are recovered, we can deduce that the completeness of \yattalens\ is comparable to that of \ringfinder.
We cannot draw more quantitative statements on completeness at this stage because 1) the SL2S lenses have been used to optimize the parameters of the algorithm and 2) the SL2S sample is not a complete sample in the first place, meaning that there could be lenses missed by \ringfinder\ that \yattalens\ would be able to identify.
In order to robustly determine the completeness of a lens search with \yattalens, we need to run the algorithm on realistic simulation of a large set of lenses.
This is left for future work.

\section{CHITAH}\label{sect:chitah}

\chitah\ \citep{Cha++15} is a lens hunter in imaging surveys, 
which is originally developed for lensed quasars, based on the configuration of lensed images. 
Not only lensed quasars, but lensed galaxies can also be captured when \chitah\ identifies lensed images within a lensed arc. 
Briefly, the procedure of \chitah\ is as follows: 
\begin{enumerate}
\item choose two image cutouts, one from bluer bands ({\it g}/{\it r}) and one from redder bands ({\it z}/{\it y}) which have sharper PSFs. 
\item match PSFs in the two selected bands. 
\item disentangle lens light and lensed arc image according to color information, producing a ``lens'' image and a ``lensed arc'' image. 
\item identify lens center and lensed image positions, masking out the region within $0.5''$ in radius from the lens center in the lensed arc image 
to prevent misidentifying lensed image positions near the lens center due to imperfect lens light separation. 
\item model the lensed image configuration with an SIE lens mass distribution. 
\end{enumerate}
The outputs of the model are the best-fit parameters of the SIE: 
the Einstein radius ($\theta_{\rm E}$), the axis ratio ($q$), the position angle (PA), the lens center, 
and the $\chi^2_{\rm src}$ on the source plane, which is defined as
\begin{equation}
\label{equ:chi2src}
\chi^2_{\rm src}= \sum_k{|{\bf r}_k-{\bf r}_{\mathrm{model}}|^2\over\sigma_{\mathrm{image}}^2/\mu_k},
\end{equation}
where ${\bf r}_k$ is the respective source position mapped from the
position of lensed image $k$ identified in the lensed arc image, $\mu_k$ is the
magnification at the position of lensed image $k$, 
$\sigma_{\mathrm{image}}$ is chosen to be the pixel scale of HSC ($0.168''$) as an estimate of the uncertainty, 
and ${\bf r}_{\mathrm{model}}$ is the modeled source position evaluated as a weighted mean of ${\bf r}_k$,
\begin{equation}
\label{equ:src_mod}
{\bf r}_{\mathrm{model}} = {\sum\limits_k{\sqrt{\mu_k}{\bf r}_k} \over \sum\limits_k{\sqrt{\mu_k}}}
\end{equation}
{\citep{Ogu10}. Here the index $k$ runs from 1 to 4 for quad systems. 
We also use the lens center from the light profile as a prior to constrain the center of the SIE lens mass model.
Therefore, we define the $\chi_{\mathrm{c}}^2$ as
\begin{equation}
\label{equ:chi2c}
\chi_{\mathrm{c}}^2= { |{\bf x}_{\mathrm{model}}-{\bf x}_{\mathrm{c}}|^2\over\sigma_{\mathrm{c}}^2},
\end{equation}
where ${\bf x}_{\mathrm{c}}$ is the lens center from the light profile,
and ${\bf x}_{\mathrm{model}}$ is the the lens center of the SIE model, 
We choose $\sigma_{\mathrm{c}}$ to be the same as $\sigma_{\mathrm{image}}$.  
We further take into account the residuals of the fit to the ``lensed arc'' image from \chitah.
The difference between the lensed arc image $Q(i,j)$ and the predicted image $Q^{\rm P}(i,j)$ is defined as,
\begin{equation}
\label{equ:chi2res}
\chi_{\mathrm{res}}^2 = \sum_{i,j} \frac{{[Q(i,j)-Q^{\rm P}(i,j)]^2}}{var(i,j)},
\end{equation}
where $i=1...N_{\rm x}$ and $j=1...N_{\rm y}$ are the pixel 
indices in the image cutout of dimensions $N_{\rm x}\times N_{\rm y}$, and $var(i,j)$ is the pixel uncertainty in $Q(i,j)$.
Each image cutout is $7\arcsec \times 7\arcsec$.

The criteria of classification of lens candidates are $\chi^2_{\rm src}+\chi_{\mathrm{c}}^2 < 2 \theta_{\rm E}$, 
where $\theta_{\rm E}$ is measured in arcsec, and $\chi_{\mathrm{res}}^2 < 100$.
The former criterion allows \chitah\ to detect lens candidates covering a wide range of $\theta_{\rm E}$, 
since typically $\chi^2_{\rm src}$ scales with $\theta_{\rm E}$ 
and our tests with mock systems in \citet{Cha++15} show that $\chi^2_{\rm src} \lesssim 4$ yields a low false positive rate of $<3\%$.
The latter criterion allows us to further eliminate false positives.
The lens candidates are selected within $0.3\arcsec < \theta_{\rm E} < 4\arcsec$.

\section{Spectroscopic search}\label{sect:emline}
Another powerful and efficient lens search method is the spectroscopic selection technique developed by \citet{Bolton04}. This spectroscopic selection technique has lead to discoveries of almost 200 strong lenses in several dedicated lens surveys including the Sloan Lens ACS Survey (SLACS, \cite{Bolton08}), the Sloan WFC Edge-on Late-type Lens Survey (SWELLS, \cite{Treu11}), the SLACS for the Masses Survey (S4TM \cite{Shu15}), the BOSS Emission-Line Lens Survey (BELLS, \cite{Brownstein12}), and the BELLS for GAlaxy-Ly$\alpha$ EmitteR sYstems Survey (BELLS GALLERY, \cite{Shu16a}, \cite{Shu16b}). Here we describe briefly the spectroscopic selection algorithm implemented in this work. More technical details can be found in \citet{Bolton04} and \citet{Shu16a}. 

For each observed spectrum, the best-fit galaxy template from the BOSS data reduction pipeline \citep{Bol++12b} is subtracted to obtain its residual spectrum. An error-weighted matched-filter search for emission-line features is performed on the residual spectrum. Note that flux errors are rescaled in this step because the BOSS pipeline-reported errors are usually underestimated at wavelengths with strong airglow lines. The rescaling process is detailed in \citet{Shu16a}. Emission-line detections with signal-to-noise ratio (SNR) greater than 4 are retained and referred to as ``hits''. In previous lens surveys, galaxies with either multiple hits (SLACS, SWELLS, S4TM, BELLS) or a single hit (BELLS GALLERY) are targeted because they fall into different lens categories. The background sources are star-forming galaxies for multiple-hit systems, and Ly$\alpha$ emitters for single-hit systems. However, as the purpose of this work is to find as many strong lenses as possible, we keep all the targets with at least one hit as lens candidates. Further visual inspections on their HSC images will confirm the lens nature. 

\section{Results}\label{sect:results}

We ran \yattalens, \chitah, and the spectroscopic search method on a sample of $\ndata$ massive galaxies from the BOSS survey.
A first set of $\sim8000$ objects was used to further optimize \yattalens, in particular to improve its purity, i.e. to reduce the number of false positives.
Once the algorithm was stable we applied it to the full sample.
\yattalens\ found $\nyatta$ lens candidates, of which 250 from the LOWZ subsample and 1230 from CMASS.
\chitah\ found 819, while the spectroscopic method found 233.
Of these candidates, 118 were common to at least two methods and 3 were found by all three.

Ten of these candidates are known lenses. They are six galaxy scale SL2S lenses (SL2SJ021411$-$040502, SL2SJ021737$-$051329, SL2SJ022346$-$053418, SL2SJ022511$-$045433, SL2SJ023307$-$043838, SL2SJ220506$+$014703), the ``Eye of Horus'' lens HSCJ142449$-$005321, already discussed in subsection \ref{ssec:known}, as well as three group scale lenses from the Strong Lensing Legacy Survey - ARCS (SARCS) sample (SL2SJ020929$-$064312, SL2SJ021408$-$053530, SL2SJ221418$+$011036, \cite{Mor++12}).
All of these were identified by \yattalens\ and two of them were found by \chitah\ as well.

We visually inspected the remaining lens candidates and graded them according to their likelihood of being strong lenses, $\plens$, using the following scheme:
\begin{itemize}
\item Grade A: definite lenses ($\plens > 0.997$), 
\item Grade B: probable lenses ($0.5 < \plens < 0.997$)
\item Grade C: possible lenses ($0.003 < \plens < 0.5$)
\item Grade 0: non lenses ($\plens < 0.003$)
\end{itemize}
A first classification was done using the image configuration as the only criterion to establish the likelihood of a candidate of being a lens.
Typical aspects that are taken in consideration are: the curvature and orientation of the candidate arc, the presence of counter-images in the opposite position to the main arc with respect to the lens, the presence of spiral arms or a disk component (suggesting that the system is not a lens).
Nine people independently graded each candidate, assigning an integer score between 0 and 3 (0 corresponding to Grade 0, 1 to Grade C, 2 to Grade B and 3 to Grade A).
A first list of grade A, B and C systems is compiled, taking the average score among the graders, rounded to the nearest integer (i.e. systems with an average score strictly larger than $2.5$ are temporarily labeled as Grade A).

We then visually inspected the spectra of the grade A, B and C candidates with emission line detections, as well as the grade A and B candidates from \yattalens\ and \chitah, to determine whether these detections are robust and, if so, to measure the redshift of the candidate lensed source.
Since we use a relatively low S/N threshold in the emission line detection algorithm, most of the candidate lines could not be confirmed by eye. 
The few systems with unambiguous line detections showed only one line.
Roughly half of these lines showed a double peak profile typical of the [OII] doublet at 3727\AA. We measured their redshift based on this interpretation.
The remaining lines were all detected at a wavelength bluer than 5100\AA. In those cases we assumed the line to be Ly-$\alpha$.

We found eight objects with visually confirmed emission lines from an object at a higher redshift of the source, all of which were already part of the emission line selected sample.
We then discussed the grade A, B and C list in light of this additional information.
Candidates HSCJ085855$-$010208 and HSCJ141815$+$015832 were upgraded from B to A, while HSCJ144307$-$004056 was upgraded from C to B.
In making the final grade A list we added a unanimity requirement: all people taking part in the grading agree on the lens nature of grade A systems.
\begin{table}
 \tbl{Lens candidate statistics.}
{
 \begin{tabular}{lcccc}
 \hline
  & \yattalens\ & \chitah\ & Emission line & Total \\
\hline
Candidates & 1480 & 819 & 233 & 2411 \\
Grade A & 15 & 8 & 3 & 15 \\
Grade B & 31 & 10 & 3 & 36 \\
Grade C & 217 & 39 & 49 & 282 \\
Known & 10 & 2 & 0 & 10 \\

 \hline
 \end{tabular}
}
\label{tab:summary}
 \begin{tabnote}{Number of lens candidates found by each search method. The first row lists the number of candidates that have been visually inspected. The second to fourth row list the number of grade A, B and C lenses respectively. The fifth row lists the number of previously known lenses that have been recovered.}
\end{tabnote}
\end{table}

\Tref{tab:summary} lists the number of candidates of each grade found by each method.
{\em We found $\ngradeA$ new grade A (definite) lenses, $\ngradeB$ grade B (probable) lenses and $\ngradeC$ grade C (possible) lenses}.
Grade A lenses are shown \Fref{fig:gradeA} while grade B candidates are shown in \Fref{fig:gradeB}. 
Their basic properties are listed in \Tref{tab:gradeAB}. The full list of candidates including grade C systems can be found online\footnote{http://hsc.mtk.nao.ac.jp/ssp/science/strong-lensing}.
\begin{figure*}
 \includegraphics[width=\textwidth]{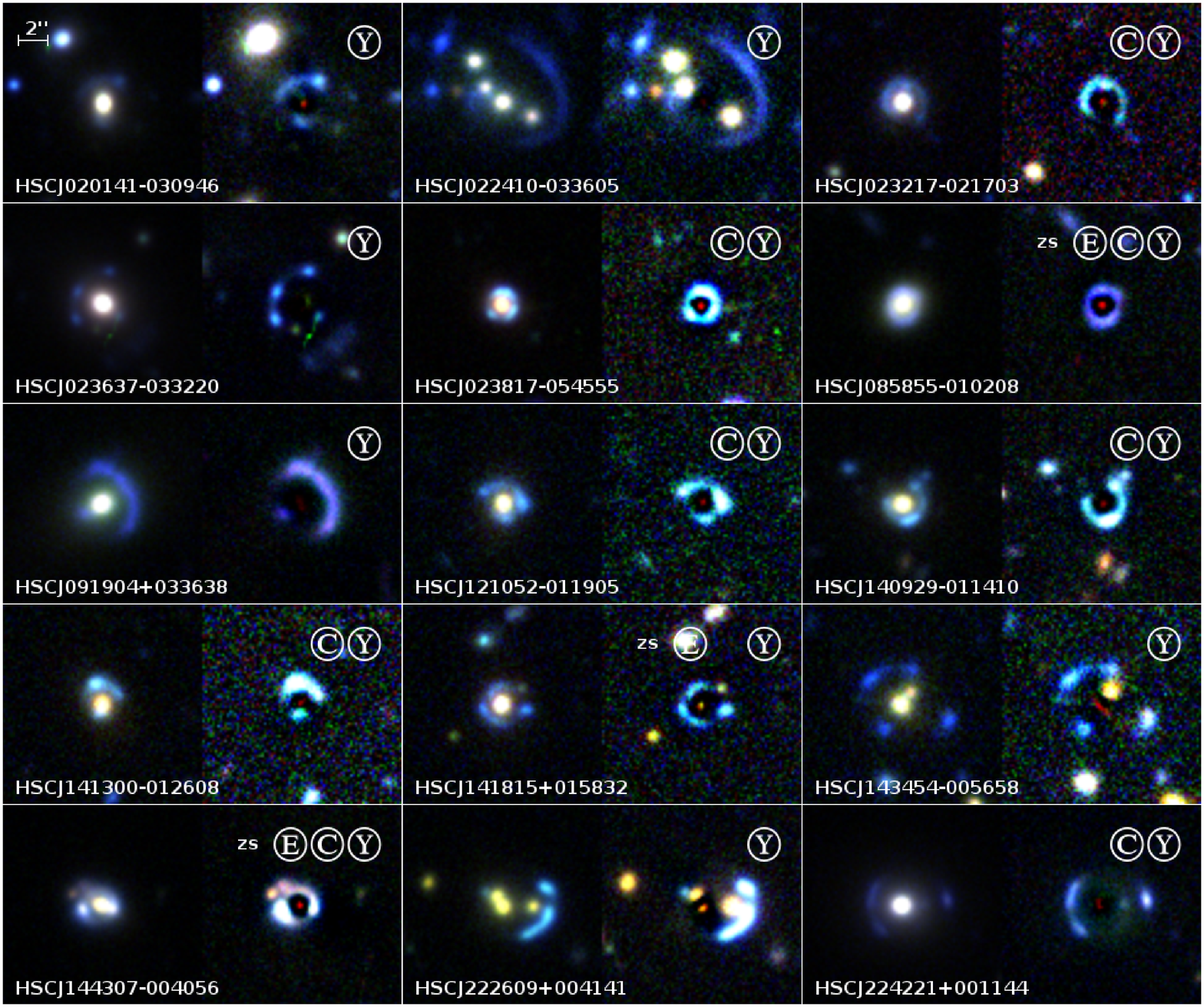}
 \caption{Grade A lenses. For each system, the left panel is a color-composite image in $g$, $r$ and $i$ bands, while the right panel is a lens-subtracted version of the image. A higher contrast is used in the right panel to enhance the images of the lensed source. Circled letters on the top of each image indicate whether the lens was found by \yattalens\ (Y), \chitah\ (C) or the emission line search (E). The letters zs indicate systems for which a spectroscopic redshift of the source galaxy has been measured. \label{fig:gradeA}
}
\end{figure*}
\begin{figure*}
\includegraphics[width=\textwidth]{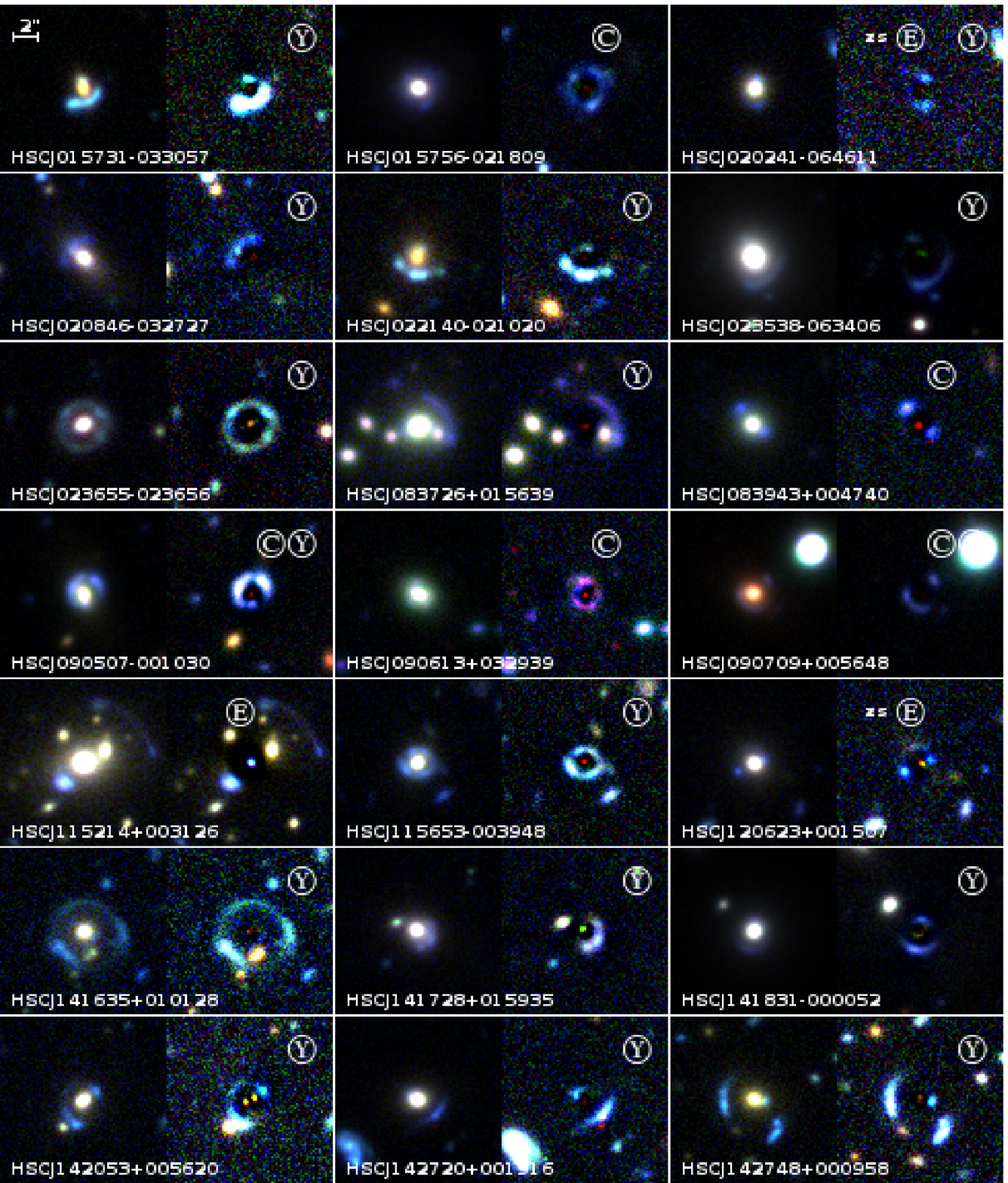}
 \caption{Grade B candidates. For each system, the left panel is a color-composite image in $g$, $r$ and $i$ bands, while the right panel is a lens-subtracted version of the image. A higher contrast is used in the right panel to enhance the images of the lensed source. Circled letters on the top of each image indicate whether the lens candidate was found by \yattalens\ (Y), \chitah\ (C) or the emission line search (E). The letters zs indicate systems for which a spectroscopic redshift of the source galaxy has been measured.
 \label{fig:gradeB}
}
\end{figure*}
\begin{figure*}
\includegraphics[width=\textwidth]{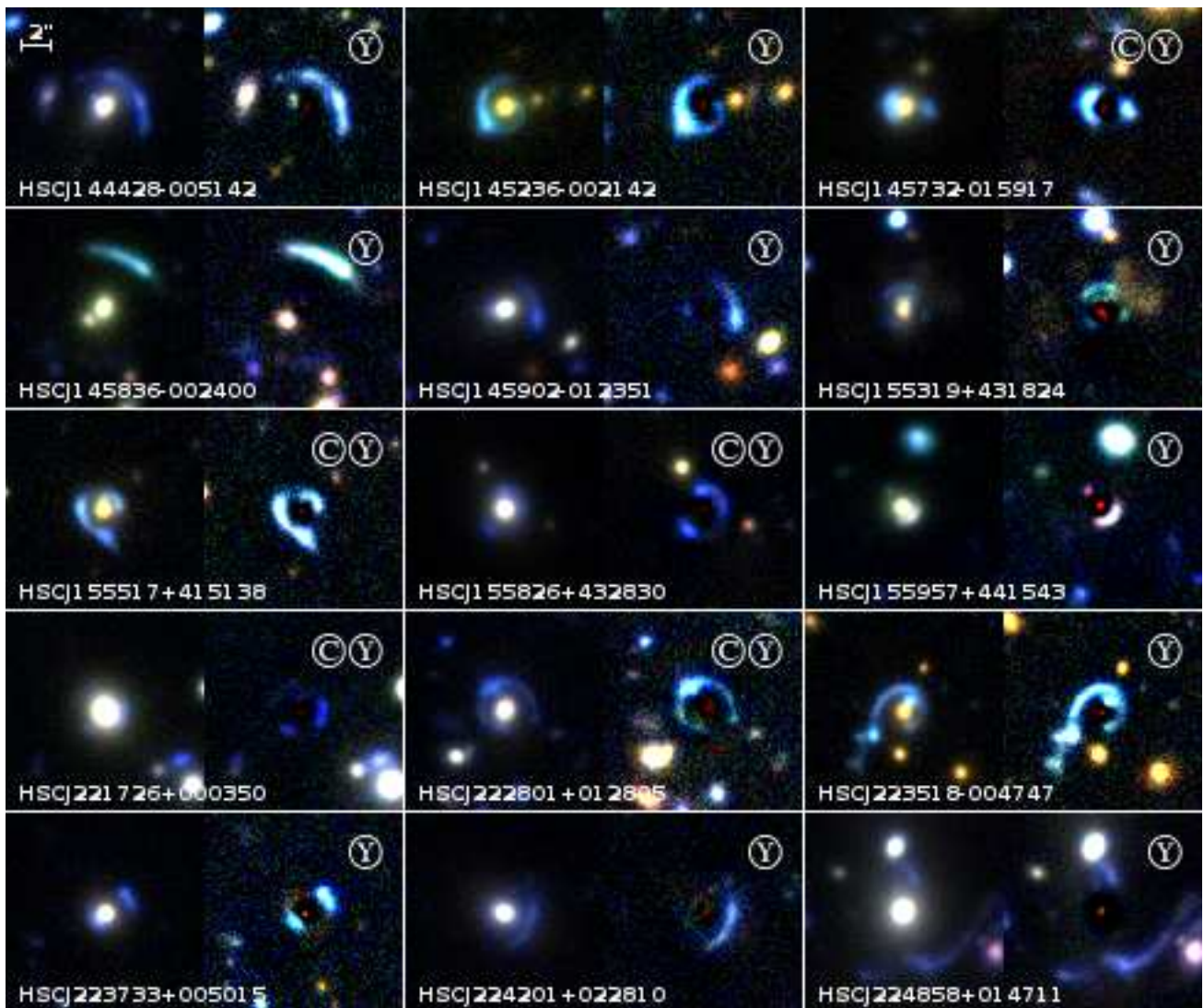}
 \caption{Grade B candidates: continued.}
\end{figure*}
\begin{table*}
 \tbl{Grade A lenses and grade B candidates.}
{
 \begin{tabular}{lcccccccccc}
 \hline
 Name & Right Ascension & Declination & $z_d$ & $z_s$ & Subsample & PDR1 & Grade & YL & EM & CH \\
 \hline
HSCJ015731$-$033057 & 01:57:31.49 & $-$03:30:57.66 & 0.621 & - & CMASS & N & B & Y & N & N \\
HSCJ015756$-$021809 & 01:57:56.61 & $-$02:18:09.96 & 0.372 & - & LOWZ & N & B & N & N & Y \\
HSCJ020141$-$030946 & 02:01:41.98 & $-$03:09:46.05 & 0.362 & - & LOWZ & N & A & Y & N & N \\
HSCJ020241$-$064611 & 02:02:41.39 & $-$06:46:11.24 & 0.502 & 2.75 & CMASS & N & B & Y & Y & N \\
HSCJ020846$-$032727 & 02:08:46.85 & $-$03:27:27.68 & 0.618 & - & CMASS & Y & B & Y & N & N \\
HSCJ022140$-$021020 & 02:21:40.13 & $-$02:10:20.10 & 0.708 & - & CMASS & N & B & Y & N & N \\
HSCJ022410$-$033605 & 02:24:10.37 & $-$03:36:05.31 & 0.613 & - & CMASS & Y & A & Y & N & N \\
HSCJ023217$-$021703 & 02:32:17.37 & $-$02:17:03.72 & 0.508 & - & CMASS & N & A & Y & N & Y \\
HSCJ023538$-$063406 & 02:35:38.22 & $-$06:34:06.07 & 0.181 & - & LOWZ & N & B & Y & N & N \\
HSCJ023637$-$033220 & 02:36:37.30 & $-$03:32:20.04 & 0.270 & - & LOWZ & N & A & Y & N & N \\
HSCJ023655$-$023656 & 02:36:55.27 & $-$02:36:56.01 & 0.562 & - & CMASS & N & B & Y & N & N \\
HSCJ023817$-$054555 & 02:38:17.77 & $-$05:45:55.52 & 0.599 & - & CMASS & N & A & Y & N & Y \\
HSCJ083726$+$015639 & 08:37:26.18 & 01:56:39.46 & 0.395 & - & LOWZ & N & B & Y & N & N \\
HSCJ083943$+$004740 & 08:39:43.03 & 00:47:40.79 & 0.621 & - & CMASS & N & B & N & N & Y \\
HSCJ085855$-$010208 & 08:58:55.99 & $-$01:02:08.42 & 0.468 & 1.42 & CMASS & N & A & Y & Y & Y \\
HSCJ090507$-$001030 & 09:05:7.35 & $-$00:10:30.03 & 0.494 & - & CMASS & Y & B & Y & N & Y \\
HSCJ090613$+$032939 & 09:06:13.14 & 03:29:39.98 & 0.617 & - & CMASS & N & B & N & N & Y \\
HSCJ090709$+$005648 & 09:07:9.70 & 00:56:48.42 & 0.478 & - & CMASS & Y & B & Y & N & Y \\
HSCJ091904$+$033638 & 09:19:4.60 & 03:36:38.65 & 0.444 & - & LOWZ & N & A & Y & N & N \\
HSCJ115214$+$003126 & 11:52:14.19 & 00:31:26.49 & 0.466 & - & CMASS & N & B & N & Y & N \\
HSCJ115653$-$003948 & 11:56:53.03 & $-$00:39:48.51 & 0.508 & - & CMASS & Y & B & Y & N & N \\
HSCJ120623$+$001507 & 12:06:23.85 & 00:15:07.15 & 0.563 & 3.12 & CMASS & N & B & N & Y & N \\
HSCJ121052$-$011905 & 12:10:52.49 & $-$01:19:05.17 & 0.700 & - & CMASS & N & A & Y & N & Y \\
HSCJ140929$-$011410 & 14:09:29.71 & $-$01:14:10.72 & 0.584 & - & CMASS & N & A & Y & N & Y \\
HSCJ141300$-$012608 & 14:13:0.07 & $-$01:26:08.16 & 0.749 & - & CMASS & N & A & Y & N & Y \\
HSCJ141635$+$010128 & 14:16:35.43 & 01:01:28.91 & 0.700 & - & CMASS & N & B & Y & N & N \\
HSCJ141728$+$015935 & 14:17:28.10 & 01:59:35.28 & 0.401 & - & LOWZ & N & B & Y & N & N \\
HSCJ141815$+$015832 & 14:18:15.73 & 01:58:32.30 & 0.556 & 2.14 & CMASS & N & A & Y & Y & N \\
HSCJ141831$-$000052 & 14:18:31.41 & $-$00:00:52.65 & 0.263 & - & LOWZ & Y & B & Y & N & N \\
HSCJ142053$+$005620 & 14:20:53.62 & 00:56:20.63 & 0.616 & - & CMASS & Y & B & Y & N & N \\
HSCJ142720$+$001916 & 14:27:20.55 & 00:19:16.11 & 0.551 & - & CMASS & Y & B & Y & N & N \\
HSCJ142748$+$000958 & 14:27:48.36 & 00:09:58.76 & 0.589 & - & CMASS & Y & B & Y & N & N \\
HSCJ143454$-$005658 & 14:34:54.40 & $-$00:56:58.56 & 0.728 & - & CMASS & Y & A & Y & N & N \\
HSCJ144307$-$004056 & 14:43:7.16 & $-$00:40:56.10 & 0.500 & 1.07 & CMASS & Y & A & Y & Y & Y \\
HSCJ144428$-$005142 & 14:44:28.74 & $-$00:51:42.45 & 0.575 & - & CMASS & N & B & Y & N & N \\
HSCJ145236$-$002142 & 14:52:36.66 & $-$00:21:42.04 & 0.733 & - & CMASS & N & B & Y & N & N \\
HSCJ145732$-$015917 & 14:57:32.58 & $-$01:59:17.36 & 0.526 & - & CMASS & N & B & Y & N & Y \\
HSCJ145836$-$002400 & 14:58:36.29 & $-$00:24:00.77 & 0.595 & - & CMASS & N & B & Y & N & N \\
HSCJ145902$-$012351 & 14:59:2.72 & $-$01:23:51.17 & 0.482 & - & CMASS & N & B & Y & N & N \\
HSCJ155319$+$431824 & 15:53:19.39 & 43:18:24.31 & 0.629 & - & CMASS & N & B & Y & N & N \\
HSCJ155517$+$415138 & 15:55:17.74 & 41:51:38.71 & 0.555 & - & CMASS & N & B & Y & N & Y \\
HSCJ155826$+$432830 & 15:58:26.66 & 43:28:30.83 & 0.444 & - & LOWZ & N & B & Y & N & Y \\
HSCJ155957$+$441543 & 15:59:57.55 & 44:15:43.81 & 0.598 & - & CMASS & N & B & Y & N & N \\
HSCJ221726$+$000350 & 22:17:26.44 & 00:03:50.33 & 0.398 & - & LOWZ & Y & B & Y & N & Y \\
HSCJ222609$+$004141 & 22:26:9.30 & 00:41:41.99 & 0.647 & - & CMASS & Y & A & Y & N & N \\
HSCJ222801$+$012805 & 22:28:1.98 & 01:28:05.74 & 0.647 & - & CMASS & Y & B & Y & N & Y \\
HSCJ223518$-$004747 & 22:35:18.31 & $-$00:47:47.30 & 0.640 & - & CMASS & N & B & Y & N & N \\
HSCJ223733$+$005015 & 22:37:33.54 & 00:50:15.78 & 0.604 & - & CMASS & Y & B & Y & N & N \\
HSCJ224201$+$022810 & 22:42:1.18 & 02:28:10.61 & 0.443 & - & LOWZ & N & B & Y & N & N \\
HSCJ224221$+$001144 & 22:42:21.58 & 00:11:44.71 & 0.385 & - & LOWZ & Y & A & Y & N & Y \\
HSCJ224858$+$014711 & 22:48:58.98 & 01:47:11.27 & 0.360 & - & LOWZ & N & B & Y & N & N \\

 \hline
 \end{tabular}
}
\label{tab:gradeAB}
 \begin{tabnote}{Column 4 and 5 list lens and source redshift (when measured) respectively. Column 6 lists which subsample of BOSS LRGs the lens galaxy belongs to. Column 7 indicates whether the system lies in the region covered by the public data release 1 or not. Column 8 indicates the grade of the candidate. Columns 9, 10 and 11 indicate if the candidate was identified by \yattalens, by the emission line search method or by \chitah.}
\end{tabnote}
\end{table*}

The 15 grade A lenses exhibit regular surface brightness profiles in their lensed sources, bright arcs and counter-images with positions and shapes expected for a smooth lensing potential, and, in three cases, emission lines from the background source.
The 36 grade B candidates, although very promising systems and most likely lenses, show features that leave room for ambiguity in their classification. 
These features are arcs that appear to extend excessively for a smooth mass distribution (for example HSCJ015731-033057, HSCJ022140-021020), arcs that appear to have too little curvature (HSCJ141635+010128, HSCJ223733+005015), the absence of visible counter-images (HSCJ020846-032727, HSCJ144428-005142), or lensed images that are simply too faint to allow for an unambiguous classification.
Higher resolution imaging data, or spectroscopic observations of the lensed arcs with a broad wavelength coverage, is needed to confirm the lens nature of these systems.

\section{Discussion}\label{sect:discuss}

We applied three different lens finding algorithms to the same photometric and spectroscopic dataset, finding $\ngradeAB$ definite/probable lenses.
This sample is large enough for us to gain valuable information on the relative efficiency of the three algorithms.
In particular we can focus on the relative completeness, defined as the number of highly probable lenses found by one algorithm divided by the number of lenses found by all three combined.

\yattalens\ has the highest relative completeness, with 46 systems identified out of $\ngradeAB$. In comparison, \chitah\ found 18 systems, while the emission line search produced 6 good candidates.
This result is not surprising because \yattalens\ was specifically developed for this study.
In contrast, \chitah\ has been optimized to look for lensed quasars. 
In this respect, it is remarkable that \chitah\ was able to find as many as 18 lenses, most of which are not lensed quasars.
The emission line search method instead relies entirely on the presence of emission lines from the lensed source in the $2.0''$ diameter optical fiber of BOSS \citep{Sme++13}.
The likelihood of a detection depends first of all on the redshift of the source (no significant emission lines fall in the optical part of the spectrum for objects in the redshift range $1.5 \lesssim z \lesssim 2.5$), but also on the Einstein radius and image configuration of the lens: lensed images too far out from the fiber do not leave a trace on the BOSS spectrum.
It is then not surprising that the emission line search missed many lenses, a good fraction of which have relatively large Einstein radius.
Nevertheless, it provided us with three highly probable lenses undetected by the other two methods.

Another important aspect to consider is purity, defined as the ratio between the number of true lenses and the total number of candidates returned by a lens finding algorithm.
Again, it is not possible with this data alone to establish the absolute purity achieved by of our search methods because for a large number of candidates, particularly the grade C candidates, we cannot determine with absolute certainty whether they are lenses or not.
However, as a proxy for purity we can consider the ratio between the number of candidates with grade B or above and the total number of candidates.
\yattalens\ achieved the highest ratio, with 3.1\% of the candidates being grade B lenses or better (3.7\% if we consider the 10 known lenses recovered), compared to 2.6\% for the emission line search method and 2.2\% for \chitah\ (2.4\% counting the two known lenses). 
Another way to interpret this result is that \yattalens\ required the least amount of visual inspection to find the same number of lenses, compared to the other methods.
In the next two subsections we will focus on some more aspects of our lens finding algorithm.

\subsection{Lenses missed by \yattalens}

\yattalens\ identified all of the new grade A lenses we discovered, but missed five grade B candidates.
Here we discuss briefly what went wrong with each of these systems.
\begin{itemize}
\item HSCJ015756$-$021809. This system was found by \chitah. Although \yattalens\ detected correctly a candidate arc around the lens galaxy, it was discarded because the ring model fit gave a better $\chi^2$ compared to the lens model fit.
\item HSCJ083943$+$004740. This lens consists of a doubly image compact source, probably a quasar, and was discovered by \chitah. The two images were correctly detected by \yattalens\ but since they are not tangentially elongated they were not classified as candidate arcs.
\item HSCJ090613$+$032939. This lens was also identified by \chitah. The main arc was detected by \yattalens\ but was then discarded after the modeling step, since the ring model produced a better fit to the data compared to the lens model.
The arc is relatively faint in the $g$-band, which is used for the modeling step. If $z$ and $i$-band are used for the modeling of the lens and arc light respectively, \yattalens\ recovers this candidate.
\item HSCJ115214$+$003126. The main candidate arc is too faint and was not detected by \sextractor.
Incidentally, although this system was found by the emission line search algorithm, the candidate arc is very faint and is located $5''$ away from the lens. It is then unlikely that a signal from the source galaxy was detected in the BOSS spectrum. Indeed, visual inspection of the spectrum did not show any convincing emission lines.
It is then possible that the presence of the arc is a fortuitous coincidence.
\item HSCJ120623$+$001507. Similarly to HSCJ083943$+$004740, this doubly imaged compact source was not recognized as a candidate arc by \yattalens\ due to the absence of any tangential elongation.
\end{itemize}

\subsection{The performance of \yattalens}

As described extensively in \Sref{sect:yatta}, \yattalens\ consists of two main steps: candidate arc detection and lens modeling.
Of the initial $\ndata$ systems, \yattalens\ detected arcs in $\narc$ of them.
Of these arcs, $\nred$ were discarded because deemed too red, while the remaining $\nmodel$ were modeled.
After the modeling, the sample size was reduced to $\nyatta$ systems, which were then visually inspected.
Of these systems, 273 were classified as possible lenses: grade C or above.

The modeling step reduced the number of candidates by more than a factor of three with respect to a selection based only on the presence of arc-like features.
This is a significant improvement in terms of human time needed to visually inspect candidates. 
Although only 20\% of the systems returned by \yattalens\ turned out to be possible lenses, this fraction is almost a factor of two larger than what was achieved by \ringfinder\ with similar data from CFHT \citep{Gav++14}.

Such an improvement in purity was accompanied by a modest loss in completeness. 
Among the 22 grade B or above lenses found by either \chitah\ or the emission line method, two were discarded by \yattalens\ during the modeling step.
Although the statistics are small to draw a robust conclusion, analysis on this sample suggests that such a loss in completeness is probably small.

\subsection{Distribution in $z-M_*$ space}

In \Fref{fig:mstarz} we plot the distribution in redshift and stellar mass of the newly found lenses, compared with the distribution of the parent sample of galaxies and of existing samples of lenses.
In particular we consider the BELLS sample, which is also a lens-selected sample based on BOSS LRGs, and the SL2S sample.
Stellar masses for the SuGOHI-g and the BELLS lenses are taken from \citet{Mar++13} and are based on spectro-photometric fitting and a Salpeter initial mass function (IMF). 
Stellar masses based on a Salpeter IMF for the SL2S lenses are taken from \citet{Son++13a}.

SuGOHI-g lenses, as well as lenses from BELLS and SL2S, are located at the high end of the mass distribution. This is expected from a lensing cross section argument: more massive objects are more likely to be strong lenses and can give rise to sets of multiple images with larger separation, thus more easily detectable.
The redshift distribution of SuGOHI-g lenses is similar to that of its parent sample, BOSS LRGs, suggesting that lensing selection and the efficiency of our lens finders do not favor strongly a particular region in redshift space.
\begin{figure}
 \begin{center}
 \includegraphics[width=\columnwidth]{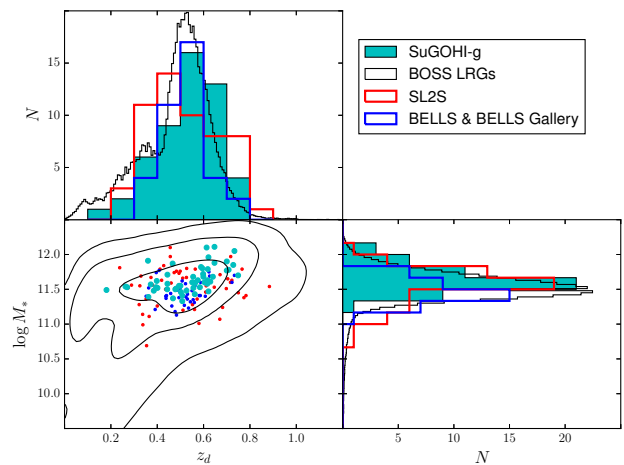}
 \end{center}
 \caption{Distribution in lens redshift and stellar mass of SuGOHI-g lenses (cyan), SL2S lenses (red), BELLS lenses (blue) and BOSS LRGs (black). The histograms of the BOSS LRGs distribution have been rescaled by an arbitrary constant.
 \label{fig:mstarz}
}
\end{figure}
\section{Conclusions and Future Prospects}\label{sect:concl}

We looked for strong gravitational lenses in the first public data release of the HSC survey.
We applied three different lens search methods to a sample of 43,000 BOSS LRGs with HSC imaging.
The first method is a new lens finding algorithm, \yattalens, developed specifically for this study. \yattalens\ looks for blue tangentially elongated features around massive galaxies, then fits a lens model to determine whether the identified features can be strongly lensed images of background galaxies.
Unlike other modeling-based lens finders, \yattalens\ does not try to determine the likelihood of an object being a lens based on the absolute quality of a lens model fit, but only in relation to alternative models. 
These models are a single S\'{e}rsic component and a profile describing the disk component of a late-type galaxy.

The second method used in the lens search is \chitah, a modeling-based algorithm originally developed to look for lensed quasars but capable of finding lensed galaxies as well.
The third method is based on the detection of emission lines from objects at a higher redshift than the main galaxy in BOSS spectroscopy data, a method that has been used successfully in various lens search campaigns.

The three methods selected a total of $\sim2,400$ candidates which were then visually inspected.
We found $\ngradeA$ new definite (grade A) lenses and $\ngradeB$ new highly probable (grade B) lenses.
\yattalens\ achieved the highest completeness and purity among the three methods with 46 grade A and B lenses out of $\nyatta$ candidates, but still missed 5 grade B lenses recovered by the other two methods.

These results are very promising in view of the advancement of the HSC SSP campaign.
The current lens search has been carried out on a 442~deg$^2$ area. 
As the HSC survey progresses to its final coverage of 1400~deg$^2$ we expect a corresponding increase in the number of lenses of more than a factor of 3.
This number is likely to be a lower limit, because $\sim60$\% of the data used in our search is shallower than the planned survey depth, potentially resulting in a decreased completeness.
Another boost in the number of lenses can be obtained by selecting targets using photometric information alone, instead of relying on BOSS spectroscopy.
For instance, the completeness of the BOSS CMASS sample at $z=0.6$ and at the mean stellar mass of the SuGOHI-g sample is 60\% \citep{Lea++16}, suggesting that selecting lens candidates using photometric redshifts could lead to an increase in sample size by almost a factor of two.
This estimate is consistent with results from the SL2S survey: of the 21 grade B or above lenses in the W1 field of CFHTLS, uniformly covered by BOSS, only 11 have BOSS spectroscopy data.
Given these considerations, since our lens finders were able to detect 61 grade B or above lenses (including the 10 known) among BOSS LRGs with HSC 2016A data, {\em we expect the SuGOHI-g sample to grow to a size of at least $\sim300$ by the end of the HSC survey and by extending the search to photometrically selected massive galaxies.}
This number is significantly more than the largest samples of lenses known to date.

\begin{ack}

The Hyper Suprime-Cam (HSC) collaboration includes the astronomical communities of Japan and Taiwan, and Princeton University.  The HSC instrumentation and software were developed by the National Astronomical Observatory of Japan (NAOJ), the Kavli Institute for the Physics and Mathematics of the Universe (Kavli IPMU), the University of Tokyo, the High Energy Accelerator Research Organization (KEK), the Academia Sinica Institute for Astronomy and Astrophysics in Taiwan (ASIAA), and Princeton University.  Funding was contributed by the FIRST program from Japanese Cabinet Office, the Ministry of Education, Culture, Sports, Science and Technology (MEXT), the Japan Society for the Promotion of Science (JSPS),  Japan Science and Technology Agency  (JST),  the Toray Science  Foundation, NAOJ, Kavli IPMU, KEK, ASIAA,  and Princeton University.

Funding for SDSS-III has been provided by the Alfred P. Sloan Foundation, the Participating Institutions, the National Science Foundation, and the U.S. Department of Energy Office of Science. The SDSS-III web site is http://www.sdss3.org/.

SDSS-III is managed by the Astrophysical Research Consortium for the Participating Institutions of the SDSS-III Collaboration including the University of Arizona, the Brazilian Participation Group, Brookhaven National Laboratory, Carnegie Mellon University, University of Florida, the French Participation Group, the German Participation Group, Harvard University, the Instituto de Astrofisica de Canarias, the Michigan State/Notre Dame/JINA Participation Group, Johns Hopkins University, Lawrence Berkeley National Laboratory, Max Planck Institute for Astrophysics, Max Planck Institute for Extraterrestrial Physics, New Mexico State University, New York University, Ohio State University, Pennsylvania State University, University of Portsmouth, Princeton University, the Spanish Participation Group, University of Tokyo, University of Utah, Vanderbilt University, University of Virginia, University of Washington, and Yale University.

This work was supported by World Premier International Research Center Initiative (WPI Initiative), MEXT, Japan.
JHHC and SHS gratefully acknowledge the support from the Max Planck Society through the Max Planck Research Group for SHS, and support from the Ministry of Science and Technology in Taiwan via grant MOST-103-2112-M-001-003-MY3.
YS is partially supported by the National Natural Science Foundation of China (NSFC) grant 11603032.
This work was supported in part by JSPS KAKENHI Grant Number 26800093 
and 15H05892.
KCW is supported by an EACOA Fellowship awarded by the East Asia Core Observatories Association, which consists of the Academia Sinica Institute of Astronomy and Astrophysics, the National Astronomical Observatory of Japan, the National Astronomical Observatories of the Chinese Academy of Sciences, and the Korea Astronomy and Space Science Institute.

\end{ack}


\end{document}